\documentclass[12pt]{article}
\usepackage{graphicx,amsfonts,amsmath,amssymb,amsthm,mathrsfs,url,hyperref}
\usepackage[usenames,dvipsnames]{xcolor}

\title{Interior-Boundary Conditions for Schr\"odinger Operators on Codimension-1 Boundaries}

\author{
Roderich Tumulka\footnote{Mathematisches Institut,
     Eberhard-Karls-Universit\"at, Auf der Morgenstelle 10, 72076
     T\"ubingen, Germany. E-mail:
     roderich.tumulka@uni-tuebingen.de}
}
\date{August 19, 2018}

\newcommand{\Hilbert}{\mathscr{H}}

\newcommand{\Q}{\mathcal{Q}}

\renewcommand{\Im}{\mathrm{Im}}

\newcommand{\RRR}{\mathbb{R}}
\newcommand{\CCC}{\mathbb{C}}

\newcommand{\SSS}{\mathbb{S}}

\newcommand{\scp}[2]{\langle #1|#2 \rangle}

\newcommand{\Laplace}{\Delta}

\newcommand{\vx}{\boldsymbol{x}}
\newcommand{\vy}{\boldsymbol{y}}

\newcommand{\vomega}{\boldsymbol{\omega}}

\newcommand{\radius}{\rho}

\newcommand{\be}{\begin{equation}}
\newcommand{\ee}{\end{equation}}

\begin{document}
\maketitle
\begin{abstract}
Interior--boundary conditions (IBCs) are boundary conditions on wave functions for Schr\"odinger equations that allow that probability can flow into (and thus be lost at) a boundary of configuration space while getting added in another part of configuration space. IBCs are of particular interest because they allow defining Hamiltonians involving particle creation and annihilation (as used in quantum field theories) without the need for renormalization or ultraviolet cut-off. For those Hamiltonians, the relevant boundary has codimension 3. In this paper, we develop (what we conjecture is) the general form of IBCs for the Laplacian operator (or Schr\"odinger operators), but we focus on the simpler case of boundaries with codimension 1.

\medskip

  \noindent 
%
\end{abstract}

\section{Introduction}
\label{sec:codim1}

Interior--boundary conditions (IBCs) are a type of boundary condition for wave functions in Schr\"odinger equations that allows the loss of probability at a boundary $\partial\Q$ due to flux into $\partial \Q$ while at the same time the probability gets added in another part of configuration space $\Q$. In contrast, ordinary boundary conditions (such as Dirichlet or Neumann boundary conditions) reflect all waves that reach $\partial \Q$, with the consequence that no probability can get lost at $\partial \Q$. IBCs come up in particular in the context of theories with particle creation and annihilation, in which probability needs to get shifted from the $n$-particle sector of configuration space to the $n+1$-particle sector and vice versa \cite{LP30,Mosh51a,Mosh51b,Tho84,Yaf92,Tum04}. IBCs have attracted interest recently for providing Hamiltonians with particle creation without the need for renormalization or ultraviolet cut-off \cite{TT15a,ibc2a,LS18,Lam18}. For a gentle introduction to IBCs, see \cite{TT15b}. Other recent works about IBCs include \cite{KS15,ST18,Sch18}.

Our goal in this paper is to develop the general form of an IBC that conserves probability (and thus can lead to a self-adjoint Hamiltonian) for Schr\"odinger operators (i.e., non-relativistic Hamiltonians whose kinetic part is the negative Laplacian) for a boundary of codimension~1. This form is similar to certain IBCs on codimension-3 boundaries considered in \cite{Yaf92,TT15a,ibc2a}. IBCs for the Dirac equation with codimension-1 boundaries are studied in \cite{IBCdiracCo1}. Particular such IBCs are used in a model of particle creation in one space dimension, involving a codimension-1 boundary, that is studied in \cite{LN18}. Preliminary considerations in the direction of this paper were described in \cite{Tum04}.

Codimension-1 boundaries form a natural framework for considering boundary conditions, including IBCs, although applications often have codimension-3 boundaries.
To put things into perspective, we mention that in the application to particle creation, the configuration space consists of configurations of a variable number of particles,
\be
\Q = \bigcup_{n=0}^\infty \Q_n = \bigcup_{n=0}^\infty \bigl( \RRR^{nd} \setminus \Delta_n \bigr)\,,
\ee
where $n$ means the particle number, $d$ the dimension of physical space, and $\Delta_n$ the set of collision configurations,
\be
\Delta_n = \bigl\{ (\vx_1,\ldots,\vx_n)\in \RRR^{nd}: \vx_i=\vx_j \text{ for some }i\neq j \bigr\}\,.
\ee
The boundary $\partial \Q$ is then $\cup_n \Delta_n$, and the IBC relates values (or limits) of $\psi$ on $\Delta_n$ to values of $\psi$ on $\Q_{n-1}$, the $n-1$-particle sector of $\Q$. In this case, the codimension of the boundary is $d$, so the physical case has codimension~3. Still, in spherical relative coordinates, the boundary corresponds to $r=0$, which in coordinates is a surface of codimension~1, and this allows us to carry over some considerations for codimension~1 to the case of codimension~3. 

This paper is organized as follows. In Section~\ref{sec:simplecodim1}, we begin by discussing a simple example. After that, we turn to the general case: the configuration and Hilbert space are set up in Section~\ref{sec:QHco1}, the IBC and Hamiltonian are written down in Section~\ref{sec:IBCco1}, and the conservation of probability is verified in Section~\ref{sec:conserveco1}. In Section~\ref{sec:radiusdelta}, we apply the IBC to an unusual kind of ultraviolet cut-off, in which the electron is smeared out not over a ball but over a sphere. In Section~\ref{sec:summary}, we summarize.

\section{Simple Example}
\label{sec:simplecodim1}

We begin with a special case that features many elements of the general discussion that will follow. Let the configuration space $\Q=\Q^{(1)}\cup \Q^{(2)}$ be the union of $\Q^{(1)}=\RRR^{d-1}$ (where $d$ is any natural number, not related to the dimension of physical space) and a half-space $\Q^{(2)}= \{(x_1,\ldots,x_d)\in\RRR^d:x_d\geq 0\}$. The boundary $\partial\Q = \partial\Q^{(2)} = \{(x_1,\ldots,x_d)\in\RRR^d:x_d=0\}$ has codimension 1. A volume measure $\mu$ on $\Q$ is defined by
\be\label{mudef1}
\mu(S)=\mathrm{vol}_{d-1}(S\cap \Q^{(1)}) + \mathrm{vol}_d(S \cap \Q^{(2)}) \,.
\ee
Wave functions are complex-valued functions on $\Q$ belonging to
\be
\Hilbert=L^2(\Q,\CCC,\mu)= L^2(\Q^{(1)})\oplus L^2(\Q^{(2)})
\ee
The IBC demands that for every $q\in\RRR^{d-1}$,
\be\label{IBCco1b}
\Bigl( \alpha(q) + \beta(q)\partial_d \Bigr)\psi(q,0) = \tfrac{2m}{\hbar^2} \, \psi(q)\,,
\ee
where $\alpha$ and $\beta$ are functions $\RRR^{d-1}\to \CCC$, and $\partial_d$ means the partial derivative with respect to the last coordinate in $\Q^{(2)}$. The Hamiltonian is defined by
\begin{align}
H\psi(q) 
&= -\tfrac{\hbar^2}{2m}\Delta\psi(q) + V(q)\,\psi(q)
+\Bigl(\gamma(q) + \delta(q)\partial_d \Bigr)\psi(q,0)
\label{Hco1bdef1}\\
H\psi(q,q_d)&=-\tfrac{\hbar^2}{2m}\Delta\psi(q,q_d) + V(q,q_d)\, \psi(q,q_d)\label{Hco1bdef2}
\end{align}
for any $q\in\RRR^{d-1}$ and any $q_d>0$. Here, $V:\Q\to\RRR$ is a potential function, and the coefficients $\gamma$ and $\delta$ (not to be confused with a Dirac delta function) are functions $\RRR^{d-1}\to\CCC$ required to satisfy the conditions
\begin{align}
&\alpha(q)^* \, \gamma(q) \in\RRR \label{acco1b}\\
&\beta(q)^* \, \delta(q) \in\RRR \label{bdco1b}\\
&\alpha(q)^* \, \delta(q) - \gamma(q)^* \, \beta(q) = -1 \label{abcdco1b}
\end{align}
at every $q\in\RRR^{d-1}$. Thus, of the 4 complex (or 8 real) degrees of freedom in the choice of coefficients $\alpha,\beta,\gamma,\delta$ at every $q$, only 4 real degrees of freedom can actually be chosen freely, while the others are determined by the conditions \eqref{acco1b}--\eqref{abcdco1b}. A slightly more restricted choice of IBCs was considered in \cite[Eq.~(25)]{TT15b}. Analogous IBCs for codimension-3 boundaries with coefficients satisfying the same relations \eqref{acco1b}--\eqref{abcdco1b} were considered in \cite{Yaf92}, \cite[Rem.~5]{TT15a}, \cite[Sec.~4]{ibc2a}. However, at codimension-3 boundaries, $\psi$ diverges like $1/r$ at the boundary, where $r$ is the distance from the boundary; in contrast, $\psi$ stays bounded at codimension-1 boundaries.

We now verify the conservation of probability on the non-rigorous level. From the Schr\"odinger equation for the Hamiltonian $H$, we obtain that for every $q\in\RRR^{d-1}$ and $q_d>0$,
\begin{align}
\frac{\partial|\psi(q)|^2}{\partial t} 
&= -\mathrm{div}\, j(q) + \tfrac{2}{\hbar} \Im \Bigl[
\psi(q)^*\bigl(\gamma(q) + \delta(q)\partial_d \bigr)\psi(q,0) \Bigr]\label{contico1b1}\\
\frac{\partial |\psi(q,q_d)|^2}{\partial t}
&=-\mathrm{div}\, j(q,q_d)\,,\label{contico1b2}
\end{align}
where $j$ means the usual probability current in either $\Q^{(1)}$ or $\Q^{(2)}$ and $\mathrm{div}$ the divergence. For each $q\in\RRR^{d-1}$, the last term in \eqref{contico1b1} can be written, by virtue of the IBC \eqref{IBCco1b}, as
\begin{align}
&\tfrac{2}{\hbar} \Im \Bigl[
\tfrac{\hbar^2}{2m}
\bigl( \alpha(q)^* + \beta(q)^*\partial_d \bigr)\psi(q,0)^*
\bigl(\gamma(q) + \delta(q)\partial_d \bigr)\psi(q,0) \Bigr]\\
&\qquad = \tfrac{\hbar}{m} \Im \Bigl[\psi(q,0)^* \, \alpha(q)^* \, \gamma(q) \, \psi(q,0) \Bigr] \nonumber\\
&\qquad + \tfrac{\hbar}{m} \Im \Bigl[\psi(q,0)^*\, \alpha(q)^* \, \delta(q) \, \partial_d\psi(q,0) \Bigr] \nonumber\\
&\qquad - \tfrac{\hbar}{m} \Im \Bigl[\psi(q,0)^* \, \gamma(q)^* \, \beta(q) \, \partial_d\psi(q,0) \Bigr] \nonumber\\
&\qquad + \tfrac{\hbar}{m} \Im \Bigl[\partial_d\psi(q,0)^* \, \beta(q)^* \, \delta(q) \, \partial_d\psi(q,0) \Bigr]\,,
\end{align}
where the first and the last term vanish by virtue of \eqref{acco1b} and \eqref{bdco1b}. What remains is
\be
\tfrac{\hbar}{m} \Im \Bigl[\psi(q,0)^* \bigl(\alpha(q)^* \, \delta(q) - \gamma(q)^* \, \beta(q) \bigr) \partial_d \psi(q,0) \Bigr]\,,
\ee 
which agrees, by virtue of \eqref{abcdco1b}, with
\be
-\tfrac{\hbar}{m} \Im\Bigl[\psi(q,0)^* \partial_d \psi(q,0) \Bigr] = - j_d(q,0) \,,
\ee
i.e., the negative of the last component of $j$ on $\partial \Q^{(2)}$. Thus, the gain in $\Q^{(1)}$ compensates the loss in $\Q^{(2)}$, so that $\|\psi\|^2 = \int_{\Q} |\psi(q)|^2 \, \mu(dq)$ is conserved.

After this simple example, we now turn to the general discussion of IBCs on codimension-1 boundaries.

\section{General Setting}
\label{sec:QHco1}

We take the configuration space $\Q$ to be a finite or countable union of disjoint manifolds with boundary, $\Q=\cup_n \Q^{(n)}$. (By definition, in a manifold with boundary, a neighborhood of an interior point looks like a piece of $\RRR^d$, while a neighborhood of a boundary point looks like a piece of a half-space in $\RRR^d$. In particular, the boundary has codimension 1, i.e., dimension $d-1$. The boundary may be empty.) We write $\partial \Q^{(n)}$ for the boundary of $\Q^{(n)}$, $\partial \Q=\cup_n \partial \Q^{(n)}$, and $\Q^\circ=\Q\setminus \partial \Q$ for the interior of $\Q$. We take $\Q$ to be equipped with a Riemann metric $g_{ij}$, which also defines a volume measure $\mu^{(n)}$ on $\Q^{(n)}$, and thus a measure $\mu$ on $\Q$,
\be\label{mudef}
\mu(S)=\sum_n \mu^{(n)}(S\cap \Q^{(n)}) \,.
\ee
Likewise, the metric defines a surface area measure $\lambda$ on $\partial \Q$. 

Wave functions can be complex-valued functions on $\Q$. However, we can also be more general and allow cross-sections of vector bundles. Readers unfamiliar with vector bundles may ignore this further generality and think of complex-valued wave functions. So, for every $n$, let $E^{(n)}$ be a \emph{Hermitian vector bundle} over $\Q^{(n)}$ of finite rank (dimension of the fiber spaces)
\be 
r_n=r(q)=\dim_{\CCC} E^{(n)}_q \,,
\ee
i.e., a complex vector bundle equipped with a positive definite Hermitian inner product $(\ , \ )_q$ in every fiber $E^{(n)}_q$, $q\in\Q^{(n)}$, and a metric connection (i.e., a connection relative to which the inner product is parallel, or, equivalently, a connection such that the parallel transport it defines along any path from $q$ to $q'$ is a unitary isomorphism $E_q^{(n)} \to E_{q'}^{(n)}$).
We write $E$ for $\cup_n E^{(n)}$ and $E_q$ for $E_q^{(n)}$ if $q\in\Q^{(n)}$. The wave function will be a cross-section of $E$, i.e., a mapping $\psi:\Q \to E$ such that $\psi(q)\in E_q$ for every $q\in\Q$. 

The Hilbert space $\Hilbert= L^2(\Q,E,\mu)$ consists of the square-integrable cross-sections of $E$ and is equipped with the inner product
\be
\scp{\psi}{\phi} = \int_{\Q} \mu(dq) \, \bigl(\psi(q),\phi(q)\bigr)_q\,.
\ee
Note that $\int_{\Q}$ means the same as $\sum_n \int_{\Q^{(n)}}$, and that $\Hilbert=\oplus_n L^2(\Q^{(n)},E^{(n)},\mu^{(n)})$.

\section{IBC and Hamiltonian}
\label{sec:IBCco1}

The IBC will be so constructed that the amount of probability per time that flows out of the boundary at $q'\in\partial \Q$ gets added to $|\psi|^2$ at an interior point
\be
q=f(q')
\ee
in a different sector, $f:\partial\Q \to \Q^\circ$. We suppose that
\be
r(f(q))\leq r(q)
\ee
(recall that $r(q)=\dim_\CCC E_q$) and further that the derivative of $f$ has full rank, i.e., that the image of $f$ in $\Q^{(n)}$ does not have lower dimension than $\Q^{(n)}$; in particular, if $q'\in\partial \Q^{(n')}$ and $q=f(q')\in\Q^{(n)}$ then $\dim \Q^{(n)}\leq \dim\partial \Q^{(n')} = \dim\Q^{(n')}-1$. 
Since many boundary points $q'$ can be mapped to the same interior point $q$, the set of which will be denoted
\be
f^{-1}(q)=\{q'\in\partial\Q: f(q')=q\}\,,
\ee
we will need to make use of a measure over $f^{-1}(q)$. The appropriate (unnormalized) measure for our purpose is
\be
\nu_q(\cdot) = \text{weak-}\!\!\!\!\lim_{dq\to \{q\}} \frac{\lambda(\,\cdot\, \cap f^{-1}(dq))}{\mu(dq)}\,,
\ee
or, equivalently, the measure characterized by
\be
\int_{\Q} \mu(dq) \, \nu_q \bigl( M\cap f^{-1}(q) \bigr) =\lambda (M)
\ee
for any $M\subseteq \partial\Q$. For example, if $S=f^{-1}(q)\cap \partial \Q^{(n')}$ is a submanifold of $\partial \Q^{(n')}$ of dimension $k$ and $\partial\Q^{(n')}$ has dimension $\ell$, then the density of $\nu_q$ relative to the volume measure arising from the Riemann metric on $S$ is
\be\label{nuqdensity}
\left|\frac{\det \Bigl(g_{\partial\Q^{(n')}} (e_i,e_j)\Bigr)_{\! i,j\leq \ell}}
{
\det \Bigl(g_{S} (e_i,e_j) \Bigr)_{\! i,j\leq k}\; 
\det \Bigl(g_{\Q^{(n)}}\bigl(df(e_i), df(e_j)\bigr)\Bigr)_{\! k<i,j\leq \ell} 
}\right|^{1/2}\,,
\ee
where $df: T_{q'}\partial\Q \to T_q\Q$ is the derivative (tangent mapping) of $f$, and $e_i$ are any linearly independent vectors in $T_{q'}\partial \Q$ with the first $k$ in $T_{q'}S$; the quantity \eqref{nuqdensity} does not depend on the choice of $e_i$.\footnote{Alternatively, $\nu_q$ can be expressed as a differential form $\hat\nu_q$ of maximal degree on $S$,
$\hat\nu_q(q')(v_1,\ldots,v_k)=\hat\lambda(q')(v_1,\ldots,v_k,e_1,\ldots, e_{\ell})$
for any $q'\in S$ and $v_1,\ldots,v_k\in T_{q'}S$, where $\hat\lambda$ is the differential form corresponding to the measure $\lambda$ (i.e., the Riemannian volume form on $\partial \Q^{(n')}$), and the $e_i$ are any vectors such that $df(e_1),\ldots, df(e_{\ell})$ is an orthonormal basis of $T_q\Q$.}
In particular, if $f^{-1}(q)$ is a finite or countable set (say, $f$ is a local diffeomorphism), then for any $q'\in f^{-1}(q)$,
\be
\nu_q(\{q'\}) = \lim_{dq'\to \{q'\}}\frac{\lambda(dq')}{\mu(f(dq'))}
=\biggl|\det \Bigl(g_{\Q^{(n)}}\bigl(df(e_i), df(e_j)\bigr)\Bigr)_{\! i,j\leq \ell}\biggr|^{-1/2}
\ee
for any orthonormal basis $\{e_i\}$ of the tangent space $T_{q'}\partial\Q$.

We now set up the Hamiltonian and IBC. We may include a potential, either as a function $V:\Q\to \RRR$ or more generally as a cross-section of $E\otimes E^*$ that is pointwise self-adjoint; here, $E_q^*$ denotes the dual space of $E_q$, so an element $V(q)$ of $E_q\otimes E^*_q$ corresponds to an endomorphism $E_q\to E_q$ that we also denote by $V(q)$, and $V$ being pointwise self-adjoint means that $V(q)$ is self-adjoint on $E_q$ relative to $(\ ,\ )_q$.

To formulate the IBC, we need an auxiliary Hermitian vector bundle $F$ on $\partial \Q$ such that $\dim_\CCC F_q = r(q)-r(f(q))$. The IBC demands that for every boundary point $q$,
\be\label{IBCco1}
\Bigl( \alpha(q) + \beta(q) \partial_n\Bigr)\psi(q) = \tfrac{2}{\hbar^2} \iota \, \psi\bigl(f(q)\bigr)\,,
\ee
where $\alpha(q)$ and $\beta(q)$ are given complex-linear mappings $E_q\to E_{f(q)} \oplus F_q$ (where $\oplus$ means orthogonal sum), $\iota$ is the inclusion $E_{f(q)}\to E_{f(q)}\oplus F_q$, $\iota(\chi) = \chi\oplus 0$, and $\partial_n$ means the normal derivative, i.e., the directional covariant derivative in the inward normal direction to the boundary at $q$ (normal in terms of the Riemann metric $g_{ij}$). Note that the IBC \eqref{IBCco1} consists of $r(q)$ equations, which is the number of components of $\psi(q)$. Of the mappings $\alpha(q)$ and $\beta(q)$ we require that
\be
[\alpha(q)|\beta(q)]  \text{ has full rank }r(q),
\ee
where the notation $[\alpha(q)|\beta(q)]$ (indicating the juxtaposition of two matrices) means the mapping $E_q\oplus E_q \to E_{f(q)}\oplus F_q$ that maps $\chi\oplus \phi$ to $\alpha(q)\chi + \beta(q)\phi$.

The Hamiltonian is, for any interior point $q$:
\be\label{Hco1def}
H\psi(q) = -\tfrac{\hbar^2}{2} \Laplace \psi(q) +V(q)\psi(q)
+ \!\!\int\limits_{f^{-1}(q)} \!\! \nu_q(dq')\, \Bigl(\gamma(q') + \delta(q') \partial_n\Bigr) \psi(q')\,.
\ee
Here, $\Laplace$ is the Laplace operator associated with the Riemannian metric of $\Q$ and the connection of $E$ (see, e.g., \cite{DGTTZ07} for a detailed definition), and the coefficients $\gamma(q')$ and $\delta(q')$ are given complex-linear mappings $E_{q'}\to E_{f(q')}$. 
The functions $\alpha,\beta,\gamma,\delta$ are required to satisfy at every $q\in\partial\Q$ the conditions
\begin{align}
&\alpha(q)^\dagger \, \iota\, \gamma(q): E_q\to E_q \text{ is self-adjoint relative to }(\ ,\ )_{q} \label{acco1}\\
&\beta(q)^\dagger \, \iota\, \delta(q): E_q \to E_q \text{ is self-adjoint relative to }(\ ,\ )_{q} \label{bdco1}\\
\label{abcdco1}
&\alpha(q)^\dagger\, \iota \,\delta(q) - \gamma(q)^\dagger\, P_{E_{f(q)}}\, \beta(q) = - I_{E_q}\,,
\end{align}
where $I_{E_q}$ means the identity operator on $E_q$, $P_{E_{f(q)}}$ the projection $E_{f(q)}\oplus F_q \to E_{f(q)}$, and $S^\dagger$ means, for any linear mapping $S: X \to Y$ between spaces with inner products, the adjoint mapping $Y\to X$, i.e., 
\be
\bigl(\chi, S^\dagger \phi \bigr)_X = \bigl( S\chi, \phi \bigr)_Y
\ee
for any $\chi \in X$ and $\phi\in Y$.

We think of the masses as incorporated into the metric $g_{ij}$, as in, e.g.,
\be
ds^2 = m_1dx_1^2+ m_1dy_1^2+m_1dz_1^2+m_2dx_2^2+m_2dy_2^2+m_2dz_2^2
\ee
for two particles of different mass in Euclidean space (see \cite{DGTTZ07} for further discussion). Then the mass need not be put into the prefactor of the Laplacian (in the Hamiltonian $H$ as in \eqref{Hco1def} above) or the gradient (in the current $j$ as in \eqref{jbundle} below).\footnote{This convention has the possibly undesirable consequence that, when different sectors correspond to different particle number, the Riemannian volume $\mu$ is weighted in different sectors with different powers of the mass (such as $m^{3n}$); however, this can easily be compensated by reweighting $\psi^{(n)}$ by a factor of $m^{-3n/2}$, which in turn requires, if $f(q)$ contains one particle less than $q$, a further factor of $m^{-3/2}$ in $\alpha,\beta,\gamma,\delta$.}

This completes the definition of the Hamiltonian.

\section{Conservation of Probability}
\label{sec:conserveco1}

Here is a formal (non-rigorous) derivation of the conservation of $|\psi|^2$, i.e., a check of self-adjointness of $H$ on the non-rigorous level. By $|\psi|^2$, we mean $|\psi(q)|^2 = (\psi(q),\psi(q))_q$, which is the density relative to $\mu$ of the probability distribution in $\Q$ associated with $\psi\in\Hilbert$ with $\|\psi\|=1$.
It evolves in general according to the balance equation
\be\label{dpsi2dtH}
\frac{\partial |\psi(q)|^2}{\partial t} = \tfrac{2}{\hbar} \, \Im \bigl( \psi(q), H\psi(q) \bigr)_q\,.
\ee
It is known (e.g., \cite{DGTTZ07}) that for $H=-\tfrac{\hbar^2}{2}\Laplace$ in a Hermitian vector bundle over a Riemannian manifold,
\be
\tfrac{2}{\hbar} \, \Im \bigl( \psi(q), H\psi(q) \bigr)_q = -\mathrm{div}\, j(q)
\ee
with the probability current vector field 
\be\label{jbundle}
j(q) = \hbar \,  \Im\,\bigl( \psi(q), \nabla\psi(q) \bigr)_q
\ee
on $\Q$.
Here, $\mathrm{div}\, j$ denotes the divergence of the vector field $j$; in coordinates, $\mathrm{div}\, j = \sum_a D_a j^a$, where $D_a$ is the covariant derivative operator arising from the Riemann metric on $\Q$. The gradient $\nabla\psi$ is the $E$-valued vector field obtained from the $E$-valued 1-form that is the covariant derivative of $\psi$ by ``raising the index'' using the Riemann metric.

Now for the Hamiltonian \eqref{Hco1def}, the balance equation \eqref{dpsi2dtH} becomes
\be\label{dpsi2dtgammadelta}
\frac{\partial |\psi(q)|^2}{\partial t} 
= -\mathrm{div}\, j(q)
+ \!\!\int\limits_{f^{-1}(q)}\!\! \nu_q(dq')\,  \tfrac{2}{\hbar} \, \Im \Bigl( \psi(q), \bigl[\gamma(q') + \delta(q') \partial_n\bigr] \psi(q')\Bigr)_q\,.
\ee
For each $q'$, the integrand can be written, by virtue of the IBC \eqref{IBCco1}, as
\begin{align}
&\tfrac{2}{\hbar} \, \Im \Bigl( \tfrac{\hbar^2}{2} P_{E_q} \bigl[ \alpha(q') + \beta(q') \partial_n\bigr]\psi(q'), \bigl[\gamma(q') + \delta(q') \partial_n\bigr] \psi(q')\Bigr)_{q} \label{expr2} \\
&\qquad =  \hbar \, \Im \Bigl( \psi(q'), \alpha(q')^\dagger \iota \gamma(q') \psi(q') \Bigr)_{q'}\nonumber\\
&\qquad+ \hbar \, \Im \Bigl( \psi(q'), \alpha(q')^\dagger \iota \delta(q') \partial_n \psi(q') \Bigr)_{q'}\nonumber\\
&\qquad-\hbar \, \Im \Bigl( \psi(q'), \gamma(q')^\dagger P_{E_q} \beta(q') \partial_n \psi(q') \Bigr)_{q'}\nonumber\\
&\qquad + \hbar \, \Im \Bigl( \partial_n\psi(q'), \beta(q')^\dagger \iota \delta(q') \partial_n \psi(q') \Bigr)_{q'} \,,
\end{align}
where the first and the last term vanish by virtue of \eqref{acco1} and \eqref{bdco1}. What remains is
\be
\hbar \, \Im \Bigl( \psi(q'), \Bigl[\alpha(q')^\dagger \iota \delta(q') -  \gamma(q')^\dagger P_{E_q} \beta(q')\Bigr] \partial_n \psi(q') \Bigr)_{q'}\,,
\ee
which agrees, by virtue of \eqref{abcdco1}, with
\be\label{jnterm}
- \hbar \, \Im \Bigl( \psi(q'), \partial_n \psi(q') \Bigr)_{q'}
=- j_n(q')\,,
\ee
where $j_n(q')$ means the component of $j(q')$ normal to the boundary, or
\be\label{jndef}
j_n=j^i\, n^j \, g_{ij}
\ee
with $n$ the inward-pointing unit normal vector to the boundary.
Thus, in total,
\be\label{balanceco1}
\frac{\partial |\psi(q)|^2}{\partial t} 
= -\mathrm{div}\, j(q)
- \!\!\int\limits_{f^{-1}(q)}\!\! \nu_q(dq')\,  j_n(q')\,.
\ee

Now, if $j_n(q')<0$ then $-j_n(q')\, \lambda(dq')\, dt$ is the amount of $|\psi|^2$ weight lost in the sector containing $q'$ due to current into the boundary region $dq'$ around $q'$ within duration $dt$. Likewise, if $j_n(q')>0$ then $j_n(q')\, \lambda(dq')\, dt$ is the amount of $|\psi|^2$ weight \emph{gained} in the sector containing $q'$ due to current coming from $dq'$ within duration $dt$. That is, $j_n(q') \, \lambda(dq')\, dt$ is the net gain, positive or negative. Now the second term on the right-hand side of \eqref{balanceco1} represents a gain in the amount of $|\psi|^2$ weight (while the $\mathrm{div}\, j$ term represents transport of $|\psi|^2$ weight within one sector); in fact, the gain in the region $dq$ around $q$ within duration $dt$ is 
\be
-  \mu(dq) \, dt \!\!\int\limits_{f^{-1}(q)}\!\! \nu_q(dq')\, j_n(q') = 
- dt \!\!\int\limits_{f^{-1}(dq)}\!\! \lambda(dq')\, j_n(q')\,.
\ee
Thus, the net gain in $dq$ exactly compensates the net loss in $f^{-1}(dq)$, and 
$\|\psi\|^2=\int_\Q |\psi(q)|^2 \mu(dq)$ is conserved.

Equation~\eqref{balanceco1} can be regarded as a transport equation for the $|\psi|^2$ weight, with two types of transport: continuous motion within a sector of $\Q$, and transport between sectors of $\Q$ (either from $q'$ to $f(q')$ or from $f(q')$ to $q'$). Equation~\eqref{balanceco1} actually is a probability transport equation for the $|\psi|^2$-distributed stochastic process in $\Q$ described in \cite{bohmibc}.

It also seems clear from the above derivation of \eqref{balanceco1} that the conditions \eqref{acco1}, \eqref{bdco1}, and \eqref{abcdco1} cannot be weakened within our scheme without losing \eqref{balanceco1} and thus the self-adjointness of $H$. (Except that \eqref{acco1}--\eqref{abcdco1} may fail on a $\lambda$-null set of boundary configurations, or $\alpha,\beta,\gamma,\delta$ may be undefined on such a set.) After all, if $\psi(q')$ and $\partial_n \psi(q')$ can be chosen independently, then the only way in which \eqref{expr2} can always be equal to \eqref{jnterm} is if \eqref{acco1}, \eqref{bdco1}, and \eqref{abcdco1} are true. Now $\psi(q')$ and $\partial_n \psi(q')$ can be chosen independently by appropriate choice of initial data for $\psi$---despite the IBC \eqref{IBCco1}, which can be satisfied by appropriate choice of $\psi(f(q'))$. To be sure, \eqref{balanceco1} can be true for all $\psi$ satisfying the IBC also if the integrals in \eqref{dpsi2dtgammadelta} and \eqref{balanceco1} agree while the integrands are not equal. For example, this happens when $\gamma=0=\delta$ (so \eqref{abcdco1} is violated), $f^{-1}(q)$ contains two boundary points, say $q'$ and $q''$, and the loss at $q'$ always compensates the gain at $q''$ and vice versa (e.g., if $j_n(q')=-j_n(q'')$ and $\nu_q(\{q'\})=\nu_q(\{q''\})$). However, such possibilities lie outside our scheme, according to which the weight lost at $q'$ is added to $f(q')$, and will not be considered here.

\section{Application: Cut-Off Radius}
\label{sec:radiusdelta}

IBCs on a codimension-1 boundary can be used for implementing an unusual kind of UV cut-off, in which the source is smeared out, but not over a ball but over a sphere of small radius $\radius>0$. 

For the sake of definiteness, let us consider a model quantum field theory in $\RRR^3$ with two kinds of particles, called $x$-particles and $y$-particles in the following, in which $x$-particles can emit and absorb $y$-particles. Suppose there is only one $x$-particle, which is fixed at the origin, whereas the $y$-particles are non-relativistic spinless bosons of mass $m_y$ and can move in $\RRR^3$; the model is adapted from the ``scalar field model'' in \cite[Chap.~12]{Schw61} and the Nelson model \cite{Nel64}, and is called ``Model 2'' in \cite{TT15a}. The configuration space is $\Q = \cup_{n=0}^\infty \RRR^{3n}$.

Let us turn to the cut-off. While an $x$-particle smeared out over a ball can emit and absorb $y$-particles anywhere within that ball, an $x$-particle smeared out over a sphere can only emit and absorb $y$-particles at a distance from its center that is exactly $\radius$. That is, a $y$-particle gets absorbed as soon as it reaches distance $\radius$ from an $x$-particle, and gets created on the sphere of radius $\radius$. We exclude the possibility that any $y$-particle is ever closer than $\radius$ to the $x$-particle. This kind of UV cut-off was first described, as far as we know, in \cite{Tum04}; we will call it a ``$\radius$-cut-off'' in the following.

When we apply this cut-off to the aforementioned model, the configuration space becomes
\be
\Q = \bigcup_{n=0}^\infty \Bigl\{ (\vy_1,\ldots,\vy_n)\in\RRR^{3n}: |\vy_j| \geq \radius \: \forall j \Bigr\}\,,
\ee
whose boundary
\be
\partial \Q = \Bigl\{ (\vy_1,\ldots,\vy_n)\in \Q:  |\vy_j|= \radius \text{ for some }j \Bigr\}
\ee
has codimension 1 almost everywhere. The Hilbert space $\Hilbert$ is the subspace of $L^2(\Q)$ of functions that are permutation invariant on each sector. Let $B_\radius$ denote the open $\radius$-ball around the origin, $B_\radius = \{\vy\in\RRR^3: |\vy|< \radius\}$. We will write $y^n$ for a configuration of $n$ $y$-particles, $y^n=(\vy_1,\ldots,\vy_n)$.

The IBC of Dirichlet type demands the following: For every $\vomega\in\SSS^2$, $n\in\{0,1,2,\ldots\}$, $y^n\in(\RRR^3\setminus B_\radius)^n$,
\be\label{IBC5a}
\psi^{(n+1)}(y^n,\radius\vomega)
= -\frac{g\, m_y}{2\pi\hbar^2\radius\sqrt{n+1}}\psi^{(n)}(y^n)\,.
\ee
The associated Hamiltonian is defined by
\begin{align}
(H\psi)^{(n)}(y^n) 
&= -\frac{\hbar^2}{2m_y} \sum_{j=1}^{n} \nabla^2_{\vy_j}\psi^{(n)}(y^n)+ nE_0 \psi^{(n)}(y^n) \nonumber\\
&\quad + \frac{g\sqrt{n+1}}{4\pi}\int\limits_{\SSS^2} d^2\vomega \,
\frac{\partial}{\partial r}\Big|_{r=\radius} \Bigl(r\psi^{(n+1)}(y^n,r\vomega) \Bigr)
\label{Hdef5a}
\end{align}
at any $y^n\in\Q\setminus\partial \Q$.

In the language of Sections~\ref{sec:QHco1}--\ref{sec:conserveco1}, $g_{ij}=m_y \delta_{ij}$, $f(y^n) = y^n \setminus B_{\radius}$, the Hermitian vector bundle $E$ is the trivial rank-1 bundle $E=\Q\times \CCC$, $\mu$ is the volume as in \eqref{mudef}, 
\be
\partial \Q^{(n)} = \bigcup_{j=1}^n (\RRR^3\setminus B_\radius)^{j-1}\times \SSS^2_\radius \times (\RRR^3\setminus B_\radius)^{n-j}\,, 
\ee
$\lambda$ is locally $\mathrm{vol}_{3(j-1)}\times \mathrm{area} \times \mathrm{vol}_{3(n-j)}$,
\be\label{f-1deltacutoff}
f^{-1}(y^n)= \bigcup_{j=1}^n \{(\vy_1,\ldots,\vy_{j-1},\radius \vomega,\vy_j,\ldots,\vy_n):\vomega\in \SSS^2\}\,,
\ee
$\nu_{y^n}(d^2\vomega) = \radius^2\, d^2\vomega$ on any of the $n$ spheres in \eqref{f-1deltacutoff}, and
\begin{align}
\alpha(y^n)&=-4\pi\radius\sqrt{n+1}/gm_y\\
\beta(y^n)&=0\\
\gamma(y^n)&=0\\
\delta(y^n)&= gm_y/4\pi\radius \sqrt{n+1}\,.
\end{align}

\section{Summary}
\label{sec:summary}

We have formulated a general version of IBCs in the non-relativistic case for boundaries of codimension~1, along with the appropriate additional term in the Hamiltonian. This formulation applies also to configuration spaces that are Riemannian manifolds with boundary, to spinor-valued wave functions, and to spin spaces that form a vector bundle. We have presented a calculation verifying that total probability is conserved. We have argued that this is the most general form of IBC unless we allow that probability lost in some part of the boundary comes out of another part of the boundary. It would be of interest to have rigorous results showing that this form of IBC is the most general one, and that it defines a self-adjoint Hamiltonian. A similar study for the Dirac equation can be found in \cite{IBCdiracCo1}. As an example, we have described a model of particle creation from a source that is neither a point nor a ball but a sphere.

\end{document}